\documentclass[12pt,preprint]{aastex}  

\shorttitle{Deuterium Fractionation in IRDC G28.34+0.06} %
\shortauthors{Chen et al.} %

\begin{document} %
\title{
Deuterium Fractionation as an Evolutionary Probe in the Infrared Dark Cloud G28.34+0.06
} %

\author{Huei-Ru Chen\altaffilmark{1,2}, Sheng-Yuan Liu\altaffilmark{2}, Yu-Nung Su\altaffilmark{2}, and Qizhou Zhang\altaffilmark{3}} %

\altaffiltext{1}{Institute of Astronomy and Department of Physics, National Tsing Hua University, Hsinchu, Taiwan; hchen@phys.nthu.edu.tw.} %
\altaffiltext{2}{Institute of Astronomy and Astrophysics, Academia Sinica, Taipei, Taiwan.} %
\altaffiltext{3}{Harvard-Smithsonian Center for Astrophysics, Cambridge, MA.} %

\begin{abstract} %
We have observed the $J=3-2$ transition of $\mathrm{N_2H^+}$ and $\mathrm{N_2D^+}$ to investigate the trend of deuterium fractionation with evolutionary stage in three selected regions in the Infrared Dark Cloud (IRDC) G28.34+0.06 with the Submillimeter Telescope (SMT) and the Submillimeter Array (SMA).   
A comprehensible enhancement of roughly 3 orders of magnitude in deuterium fractionation over the local interstellar $\mathrm{D/H}$ ratio is observed in all sources.
In particular, our sample of massive star-forming cores in G28.34+0.06 shows a moderate decreasing trend over a factor of 3 in the $N(\mathrm{N_2D^+})/N(\mathrm{N_2H^+})$ ratio with evolutionary stage, a behavior resembling what previously found in low-mass protostellar cores. 
This suggests a possible extension for the use of the $N(\mathrm{N_2D^+})/N(\mathrm{N_2H^+})$ ratio as an evolutionary tracer to high-mass protostellar candidates.
In the most evolved core, MM1, the $\mathrm{N_2H^+ \; (3-2)}$ emission appears to avoid the warm region traced by dust continuum emission and emission of $\mathrm{^{13}CO}$ sublimated from grain mantles, indicating an instant release of gas-phase CO.  
The majority of the $\mathrm{N_2H^+}$ and $\mathrm{N_2D^+}$ emission is associated with extended structures larger than $8^{\prime\prime}$ ($\sim 0.2 \; \mathrm{pc}$).  
\end{abstract} %


\section{Introduction} %
In the early evolutionary stages of star formation process, sequential depletion of molecular species on grain mantles nurtures  a peculiar low-temperature chemistry due to the removal of important gas-phase reactants, starting with sulfur-bearing species and followed by even volatile molecules such as CO (Bergin \& Tafalla 2007).
Besides $\mathrm{H_2}$, $\mathrm{N_2}$ is thought to be least affected in this condensation process and results in an enrichment of its daughter products, $\mathrm{NH_3}$ and $\mathrm{N_2H^+}$ (Bergin {et~al.} 2002).
The removal of the gas-phase $\mathrm{CO}$ also promotes ion-molecular reactions and induces a sharp increase in the abundance of deuterated molecules in dense cores (Millar, Bennett, \& Herbst 1989).
Indeed, an enhancement of $2-3$ orders of magnitude in the $\mathrm{D/H}$ ratio in star-forming cores (Crapsi {et~al.} 2005; Fontani {et~al.} 2006; Pillai {et~al.} 2007) over the local interstellar value of $1.51 \times 10^{-5}$ (Oliveira {et~al.} 2003) has been observed.
In particular, the deuterium fractionation of $\mathrm{N_2H^+}$, $D_\mathrm{frac} \equiv N(\mathrm{N_2D^+})/N(\mathrm{N_2H^+})$, in low-mass star-forming cores shows an increasing trend with dynamical age in the prestellar phase (Crapsi {et~al.} 2005) but a decreasing trend in the protostellar phase (Emprechtinger {et~al.} 2009).
Chemical models anticipate $D_\mathrm{frac}$ to be affected by a few factors such as the kinetic temperature and, for ion species, the electron abundance (Roueff {et~al.} 2005) as well as the gas-phase CO abundance (Aikawa {et~al.} 2005).
The abundance of gas-phase CO is expected to decline in the prestellar phase through molecular depletion onto grain surfaces but to rise up in the protostellar phase through sublimation of ice mantles as the envelope warms up.
The correlation between deuterium fractionation and CO depletion factor has been recognized in a compiled sample of prestellar and protostellar cores (Crapsi {et~al.} 2005; Emprechtinger {et~al.} 2009).
In a subsample of Taurus cores, a better correlation is found and leads to the speculation of external environment being influential to the evolution of a core (Crapsi {et~al.} 2005).
On the other hand, there is no evidence of a consistent behavior of $D_\mathrm{frac}$ in the case of high-mass protostellar candidates although a clear but less dramatic enhancement has been observed in a number of massive protostellar objects (Fontani {et~al.} 2006; Pillai {et~al.} 2007).
The environs of high-mass protostars may not sustain low temperature long enough to build up deuterated species as abundant as those of low-mass objects.
In this study, we investigated a possible trend in the deuterium fractionation of $\mathrm{N_2H^+}$ with dynamical age using a sample of massive star-forming cores from one single IRDC to reduce the environmental fluctuations among the selected cores.

Infrared dark clouds (IRDCs) were first discovered by Infrared Space Observatory (ISO) and the Midcourse Space Experiments (MSX) through silhouette against the bright, diffuse infrared background emission of the Galactic plane (Egan {et~al.} 1998; Carey {et~al.} 1998, 2000; Hennebelle {et~al.} 2001; Rathborne, Jackson, \&  Simon 2006). 
Because of their large mass ($M \gtrsim 10^3 M_\odot$), low temperature ($T < 20 \; \mathrm{K}$), and high density ($n_\mathrm{H_2} \gtrsim 10^5 \; \mathrm{cm^{-3}}$), IRDCs have been proposed to be in the earliest stage of massive star formation.
At a distance of $4.8 \; \mathrm{kpc}$, IRDC G28.34+0.06 (hereafter G28)  is associated with roughly $10^3 \; M_\odot$ in the infrared absorption region and contains several dense cores in different evolutionary stages (Carey {et~al.} 2000; Wang {et~al.} 2008).
We have selected three dense cores in G28 to form an evolutionary sequence: starting with MM9 in an early stage of mass collection with one weak continuum source, followed by MM4 in a stage of mass fragmentation with at least five continuum sources, and MM1 in a later stage with embedded massive protostellar objects with a total luminosity of $10^3 \; L_\odot$. 
All the selected regions are associated with water masers, indicating star-forming activities (Wang {et~al.} 2006).

Our study is enabled by the Arizona Radio Observatory (ARO) Submillimeter Telescope (SMT) and the Submillimeter Array\footnotemark[4] (SMA).
The SMT offers good sensitivity to detect weak line emission in extended structures while the SMA can preferentially image compact structures in the regions of interest.   
\footnotetext[4]{The Submillimeter Array is a joint project between the Smithsonian Astrophysical Observatory and the Academia Sinica Institute of Astronomy and Astrophysics, and is funded by the Smithsonian Institution and the Academia Sinica.}  

\section{Observations and Data Reduction} %
We observed the emissions of $\mathrm{N_2H^+ \; (3-2)}$ at $279.511780 \; \mathrm{GHz}$ and $\mathrm{N_2D^+ \; (3-2)}$ at $231.321864 \; \mathrm{GHz}$ towards the three selected regions in G28 with both the SMT and SMA.

\subsection{SMT Observations} %
Single-dish observations of $\mathrm{N_2H^+ \; (3-2)}$ and $\mathrm{N_2D^+ \; (3-2)}$ toward G28-MM1, MM4, and MM9 were carried out on 2008 April 13 with the 10-meter SMT on Mount Graham, Arizona. 
The observations were performed in the beam-switching mode. 
The pointing centers followed the peak positions given by Rathborne {et~al.} (2006): ($\alpha$, $\delta$)(J2000) = ($\mathrm{18^h42^m52.10^s,-3^\circ59^\prime45.0^{\prime\prime}}$) for MM1, ($\alpha$, $\delta$)(J2000) = ($\mathrm{18^h42^m50.70^s,-4^\circ3^\prime15.0^{\prime\prime}}$) for MM4, and ($\alpha$, $\delta$)(J2000) = ($\mathrm{18^h42^m46.70^s,-4^\circ4^\prime8.0^{\prime\prime}}$) for MM9. 
The primary beam is about 26\arcsec\ for $\mathrm{N_2H^+}$ and 32\arcsec\ for $\mathrm{N_2D^+}$. 
The spectral resolution is $1 \; \mathrm{MHz}$, corresponding to a velocity resolution of $1.07 \; \mathrm{km \, s^{-1}}$ for $\mathrm{N_2H^+}$ and $1.30 \; \mathrm{km \, s^{-1}}$ for $\mathrm{N_2D^+}$. 
The temperature scale $T_\mathrm{A}^*$ was obtained using standard vane calibration, and the main beam temperature, $T_\mathrm{mb}$, was derived through $T_\mathrm{mb} = T_\mathrm{A}^*/\eta_\mathrm{mb}$ with a main beam efficiency $\eta_\mathrm{mb} = 0.75$. 
The respective rms noise level is about $T_\mathrm{mb} = 20$ and $10 \; \mathrm{mK}$ for the $\mathrm{N_2H^+}$ and $\mathrm{N_2D^+}$ data, respectively. 
Data reduction was done with the CLASS package.

\subsection{SMA Observations} %
Observations with the SMA were carried out with seven antennas in the compact configuration on 2008 June 17 for $\mathrm{N_2H^+}$ and September 30 for $\mathrm{N_2D^+}$.
The system temperature varied from 150 to 350~K during the $\mathrm{N_2H^+}$ observation and from 100 to 180~K during the $\mathrm{N_2D^+}$ observation.  
The correlator was set to have a spectral resolution of $410 \; \mathrm{kHz}$, equivalent to $0.44$ and $0.53 \; \mathrm{km \, s^{-1}}$ for $\mathrm{N_2H^+}$ and $\mathrm{N_2D^+}$, respectively.
The projected baselines ranged from $11-118\; \mathrm{k\lambda}$ and $11-59 \; \mathrm{k\lambda}$, which are insensitive to structures larger than 8\arcsec\ (Wilner \& Welch 1994). 
The full-width at half power (FWHP) of the primary beam is roughly 44\arcsec\ for $\mathrm{N_2H^+}$ and 53\arcsec\ for $\mathrm{N_2D^+}$.

The observing cycle comprised scans of 1751+096, MM1, MM4, MM9, and 1911$-$201, and was repeated every 27 minutes.
The phase centers of the first observation run were the same as those of the SMT observations.  
Based on the results of the first observation run, we adjusted the phase center for the second observation run on the continuum peak in MM1 to ($\alpha$, $\delta$)(J2000) = ($\mathrm{18^h42^m52.00^s,-3^\circ59^\prime53.00^{\prime\prime}}$) and in MM9 to ($\alpha$, $\delta$)(J2000) = ($\mathrm{18^h42^m46.46^s,-4^\circ4^\prime15.1^{\prime\prime}}$).
Data inspection, bandpass and flux calibrations, as well as temporal gain derivation were done within the IDL superset MIR.  
The flux scale was derived by observations of Uranus and is estimated to be accurate within 15\%.

Imaging was performed using MIRIAD package with natural weighting.
To keep comparison easy, we present all the maps with center positions consistent with the phase centers of the second observation run.
For every source in each observation run, we used visibilities from both sidebands, each of a 2~GHz bandwidth, to generate a line-free continuum map. 
Channel maps were made with visibilities gridded into a velocity resolution of $1 \; \mathrm{km \, s^{-1}}$ and resulted in rms noises, $\sigma = 0.31$ and $0.20 \; \mathrm{K}$, equivalent to $110$ and $82 \; \mathrm{mJy \, beam^{-1}}$, with angular resolutions of $3.3\arcsec \times 1.7\arcsec$ and $4.0\arcsec \times 2.4\arcsec$ for $\mathrm{N_2H^+}$ and $\mathrm{N_2D^+}$, respectively.
 
\section{Results and Discussion} %
\subsection{Deuterium Fractionation as an Evolutionary Probe} %

\begin{figure}
\plotone{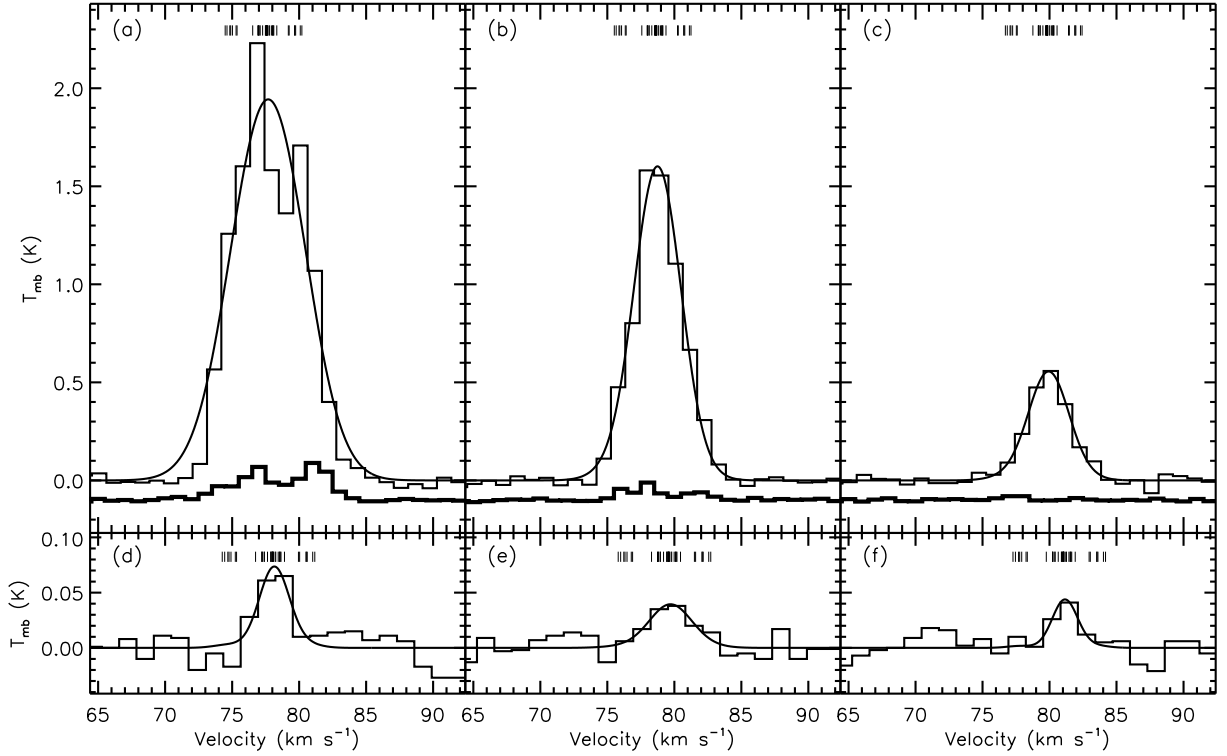}
\caption{(a) through (c) SMT $\mathrm{N_2H^+ \; (3-2)}$ spectra ({\it histogram}) towards MM1, MM4, and MM9, respectively, superposed with the thirty-eight hyperfine component model-fit spectra ({\it solid curves}) assuming a single excitation temperature.  The frequency of each individual hyperfine component is labelled with a short bar on the top.  The SMA spectra after convolving with the SMT beam are plotted with a displacement of $-0.1 \; \mathrm{K}$ ({\it thick histogram}).  (c) through (f) SMT $\mathrm{N_2D^+ \; (3-2)}$ spectra ({\it histogram}) towards MM1, MM4, and MM9, respectively, superposed with the thirty-eight hyperfine component model-fit spectra ({\it solid curves}).  \label{fig_g28smt}}
\end{figure}

The $J=3-2$ transition of $\mathrm{N_2H^+}$ and $\mathrm{N_2D^+}$ were detected in all three sources with the SMT (Fig.~\ref{fig_g28smt}).
Since both transitions contain numerous hyperfine components, we fit each spectrum of every source with a model comprised of thirty-eight hyperfine components with updated line frequencies and spontaneous emission rates (Pagani, Daniel, \& Dubernet 2009).
For each individual source, all the hyperfine components of every $J$-level are assumed to be in thermal equilibrium at a single excitation temperature, $T_\mathrm{ex}$, adopted from the ammonia observations with an angular resolution of 40\arcsec\ (Pillai {et~al.} 2006).
The models are described by three more parameters: total column density, $N$, systemic velocity, $\upsilon_\mathrm{LSR}$, full-width at half maximum (FWHM) as line width, $\Delta \upsilon$.
Model spectra are optimized with the minimization of the reduced $\chi^2$ value, $\overline{\chi^2}$, and the results are listed in Table~\ref{table_smtfits}.
The $\mathrm{N_2H^+}$ spectrum towards MM1 appears to be doubly peaked, possibly affected by the presence of multiple sources (Zhang {et~al.} 2009) and resulted in a significantly broader line width and a large $\overline{\chi^2}$. 
\placefigure{f1} %
\placetable{table_smtfits} %

In cold clouds, a correction of the cosmic background temperature, $T_\mathrm{bg} = 2.7 \; \mathrm{K}$, is necessary when extracting the optical depth information of an observed spectrum with 
\begin{equation} %
  \tau(\upsilon) = -\ln \left[ 1 - \frac{T_\mathrm{mb}(\upsilon)}{J(T_\mathrm{ex}) - J(T_\mathrm{bg})} \right], 
\end{equation} %
where $T_\mathrm{mb}(\upsilon)$ is the main beam temperature of the spectra, and $J(T) = (h\nu/k)(e^{h\nu/kT}-1)$.
All the line emissions appear to have fairly small optical depths.
Meanwhile, the optimized model also provides an estimate of optical depth by integrating optical depths of all the hyperfine components.
The maximum optical depth, $\tau_\mathrm{max}$, is less than 0.21 (Table~\ref{table_smtfits}).
The emission of all the observed lines is optically thin. 
Over all, the fitted line widths, $\Delta \upsilon$, of all the transitions are much broader than their thermal line width, $\Delta \upsilon_\mathrm{th} \equiv \sqrt{8 \ln 2 \, k T_\mathrm{ex}/m_\mathrm{N2H^+}} \simeq 0.16 \; \mathrm{km \, s^{-1}}$, suggesting a significant contribution from nonthermal motions.
Our $\mathrm{N_2H^+ \; (3-2)}$ spectra also have larger line widths when compared to the $\mathrm{NH_3 \; (1,1)}$ spectra observed with the Very Large Array (VLA) and the Effelsberg 100m telescope (Pillai {et~al.} 2007; Wang {et~al.} 2008).
Since the two transitions have similar upper state energies, $E_\mathrm{up} = 26.8 \; \mathrm{K}$ for  $\mathrm{N_2H^+}$ and $23.4 \; \mathrm{K}$ for $\mathrm{NH_3}$, but fairly different critical densities, $n_\mathrm{crit} \sim 10^6 \; \mathrm{cm^{-3}}$ for $\mathrm{N_2H^+}$ (Daniel {et~al.} 2005; Pagani {et~al.} 2009) and $10^3 \; \mathrm{cm^{-3}}$ for $\mathrm{NH_3}$ (Evans 1999), our $\mathrm{N_2H^+}$ observations tend to trace denser clumps possibly embedded in the inner region that is affected by star-forming activities as suggested by  $\mathrm{H_2O}$ masers (Wang {et~al.} 2006).
Observations of lower $J$ transitions of $\mathrm{N_2H^+}$ such as the $J=1-0$ ($E_\mathrm{up} = 4.5 \; \mathrm{K}$ and $n_\mathrm{crit} \sim 10^5 \; \mathrm{cm^{-3}}$) line are needed to verify
this interpretation.

The deuterium fractionation, $D_\mathrm{frac}$, is calculated for every source (Table~\ref{table_smtfits}) and shows a significant enhancement of roughly 3 orders of magnitude higher than the average $\mathrm{D/H}$ ratio of $1.51 \times 10^{-5}$ in the local interstellar medium (Oliveira {et~al.} 2003).
Such an enhancement has been observed in a large sample of high-mass star-forming cores (Fontani {et~al.} 2006) as well as low-mass prestellar and protostellar cores (Crapsi {et~al.} 2005; Roberts \& Millar 2007; Emprechtinger {et~al.} 2009).
In particular, we find a decreasing trend with evolutionary stage, from $D_\mathrm{frac} = 0.051$ in the younger MM9 core to $D_\mathrm{frac} = 0.016$ in the more evolved MM1 core, a change over a factor of roughly $3$.
Although a decreasing trend over a factor of roughly $8$ has been identified in low-mass protostellar cores (Emprechtinger {et~al.} 2009), there was no evidence for such trend in a sample of massive star-forming cores associated with different molecular clouds (Fontani {et~al.} 2006).  
Variations in thermal history, external environment, and initial chemical abundances across different molecular clouds may cause undesired confusion. 
Although our sample contains only three sources, the association in one single IRDC helps to minimize the environmental variations among the sources and to reveal the gentle drop in deuterium fractionation as the dusty envelope warms up.

\subsection{Emission Contained in Compact Structures \label{g28comp}}
\begin{figure} %
\plotone{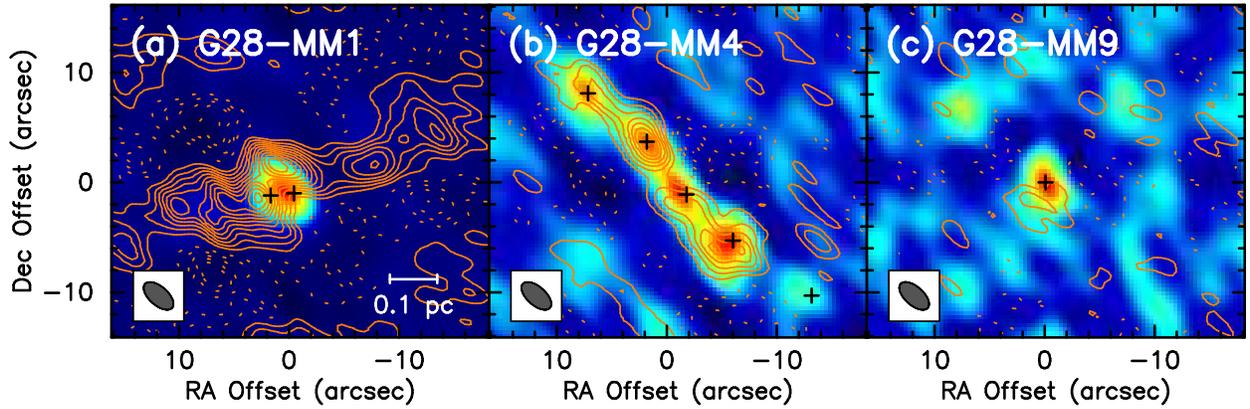} %
\caption{SMA $\mathrm{N_2H^+ \; (3-2)}$ integrated intensity ({\it contours}) overlaid on the $284.2 \; \mathrm{GHz}$ line-free continuum map towards MM1, MM4, and MM9.  
Black crosses label the continuum peaks reported by Zhang {et~al.} (2009) in MM1 and MM4 and by this work in MM9. 
Contour levels correspond to $(-10, -8, \dots, -2, 2, 4, \dots, 22) \, \sigma$, where $\sigma = 1.10, 0.93,$ and $0.88 \; \mathrm{K \, km \, s^{-1}}$ for MM1, MM4, and MM9, respectively. 
The synthesized beam shown is $3.3\arcsec \times 1.7\arcsec$ ($\mathrm{P.A.} = 49^\circ$) for the $\mathrm{N_2H^+}$ maps. 
The continuum map has an intensity range of $-48$ to $738 \; \mathrm{mJy \, beam^{-1}}$ for MM1, $-14$ to $56 \; \mathrm{mJy \, beam^{-1}}$ for MM4, and $-13$ to $46 \; \mathrm{mJy \, beam^{-1}}$ for MM9 with beam sizes of $4.0\arcsec \times 2.5\arcsec$ ($\mathrm{P.A.} = 31^\circ$).  
\label{fig_g28n2hp}} %
\end{figure}

Using the SMA, we further imaged the $\mathrm{N_2H^+}$ and $\mathrm{N_2D^+}$ emission to study more centrally concentrated structures.
The $\mathrm{N_2H^+}$ emission shows clear detections in MM1 and MM4 as well as a marginal detection in MM9 (Fig.~\ref{fig_g28n2hp}).
Since an interferometer is insensitive to structures larger than the scale corresponding to its shortest projected baseline, our SMA observations serve as a spatial filter to probe the fractional flux contained in clumps smaller than 8\arcsec\ (Wilner \& Welch 1994).
A comparison of the integrated intensity observed with the SMA, $W_\mathrm{SMA}$, to that with the SMT, $W_\mathrm{SMT}$, can be made by convolving the SMA maps with the SMT beam after masking out regions below $-3\sigma$, which are believed to be artifacts induced by the lack of short baselines.
For every source, we compare the $\mathrm{N_2H^+}$ integrated intensity observed with the SMA to that with the SMT (Fig.~\ref{fig_g28smt}) and estimate the fraction of the SMA integrated  intensity, $W_\mathrm{SMA}/W_\mathrm{SMT}$, which indicates the state of a mass concentration process (Table~\ref{table_mms}).
The small values of  $W_\mathrm{SMA}/W_\mathrm{SMT}$ suggest that most of the $\mathrm{N_2H^+}$ emission is in structures larger than 8\arcsec.
An increasing trend from MM9 to MM1 is found and agrees with the presumed evolutionary stage of the central sources.
On the other hand, $\mathrm{N_2D^+}$ is not detected in all regions at a level of $4\sigma \simeq 2.3 \; \mathrm{K \, km \, s^{-1}}$, translating to an upper limit of $N(\mathrm{N_2D^+}) = 2.0 \times 10^{12} \; \mathrm{cm^{-2}}$ at $T_\mathrm{ex} = 13 \; \mathrm{K}$.
Compared to  the SMT integrated intensity in MM9, this detection limit sets a maximum fraction of 19\% for emission coming from compact structures.    
Given the high critical density of $n_\mathrm{crit} \sim 10^6 \; \mathrm{cm^{-3}}$, the large fraction of the $\mathrm{N_2H^+} \; (3-2)$ emission missed by the SMA observations strongly suggests the presence of cold and dense gas in scales larger than 8\arcsec\ ($\sim 0.2 \; \mathrm{pc}$). 
Similar to $\mathrm{N_2H^+}$, most of the $\mathrm{N_2D^+}$ emission is in extended structures.

\placefigure{f2} %
\placetable{table_mms} %

Continuum emissions at $284.2$ and $236.5 \; \mathrm{GHz}$ are detected in all sources with similar morphology.
For the first time, a compact continuum source is detected in MM9 (Fig.~\ref{fig_g28n2hp}).
Since the visibility coverages of the two observation runs were quite different, we made continuum maps with visibilities of projected baselines within $13-57 \; \mathrm{k\lambda}$, the range that all sources have in common at the two frequencies.
If the dust emission is optically thin, the continuum flux density $F_\nu \propto \kappa_\nu \, B_\nu(T_d)$, where $\kappa_\nu = 0.006 \, (\nu/245 \; \mathrm{GHz})^{\beta} \; \mathrm{cm^2 \, g^{-1}}$ (Shepherd \& Watson 2002) is the dust opacity at the observing frequency $\nu$, and $B_\nu(T_d)$ is the Planck function at a dust temperature $T_d$.
The opacity spectral index, $\beta \equiv \Delta \log \kappa_\nu/\Delta \log \nu$, can be measured by comparing flux densities at two observing frequencies.
In cold clouds, the condition $h \nu \sim k T_d$ makes the Rayleigh-Jeans approximation inappropriate.
We estimate the opacity spectral index in the selected regions with
\begin{equation} %
    \beta = \frac{\log (F_{\nu_2}/F_{\nu_1}) + \log [(e^{h\nu_2/kT_d}-1)/(e^{h\nu_1/kT_d}-1)]}{\log (\nu_2/\nu_1)} - 3,        
\end{equation} %
and the core mass with $M_\mathrm{core} = F_{\nu} D^2/\kappa_{\nu} B_{\nu}(T_d)$, where $D$ is the distance of the source. 
Assuming thermal equilibrium between gas and dust over similar spatial scales in the calculations of $\beta$ and $M_\mathrm{core}$, we adopt the gas temperature derived from the VLA ammonia observations with a comparable angular resolution of $5\arcsec \times 3\arcsec$ (Zhang {et~al.} 2009) to be the dust temperature, $T_d$.  
The results are listed in Table~\ref{table_mms}.
Because of the limited frequency span of $48 \; \mathrm{GHz}$, we note that a systematic uncertainty of 15\% in flux density measurements will produce a fairly large uncertainty of $\Delta \beta \simeq 1.2$ in our calculations.

The dust opacity spectral index shows an increasing trend from MM9 to MM1 as the central sources evolve although the multiplicity in MM4 may complicate the interpretation of its averaged $\beta$.
Nevertheless, a smaller value of $\beta$ is observed in MM9 with respect to MM1.
At millimeter wavelengths, the value of $\beta$ is a good probe for the size distribution of dust grains (Miyake \& Nakagawa 1993), and a smaller $\beta$ suggests that MM9 is surrounded by larger dust grains as a result of grain growth in high-density environment.
On the other hand, the larger $\beta$ in the more evolved region MM1 may be attributed to possible changes in the size distribution and chemical composition of the surrounding dust grains, which have been exposed to strong radiation fields generated by the associated massive YSOs. 

\subsection{Deficiency of N$_2$H$^+$ in a Warm Region} %
In cold, dense environment, CO tends to freeze out onto dust grains while $\mathrm{N_2H^+}$ and $\mathrm{NH_3}$ suffer less from depletion (Bergin {et~al.} 2002; Tafalla {et~al.} 2004).
Since CO is the major destroyer of molecular ions, its removal from the gas phase results in a subsequent enrichment of $\mathrm{N_2H^+}$ (Aikawa {et~al.} 2005). 
In later evolutionary stages when the internal heating of YSOs becomes important, the temperature will rise up and lead to sublimation of ice mantle, which returns volatile species such as CO back to the gas phase. 
The warmer temperature together with the reappearance of gas-phase CO can alter the competition among chemical reaction routes and destroy $\mathrm{N_2H^+}$ that has been produced during the cold early phase (Roueff {et~al.} 2005; Aikawa {et~al.} 2005).

\begin{figure} %
\plotone{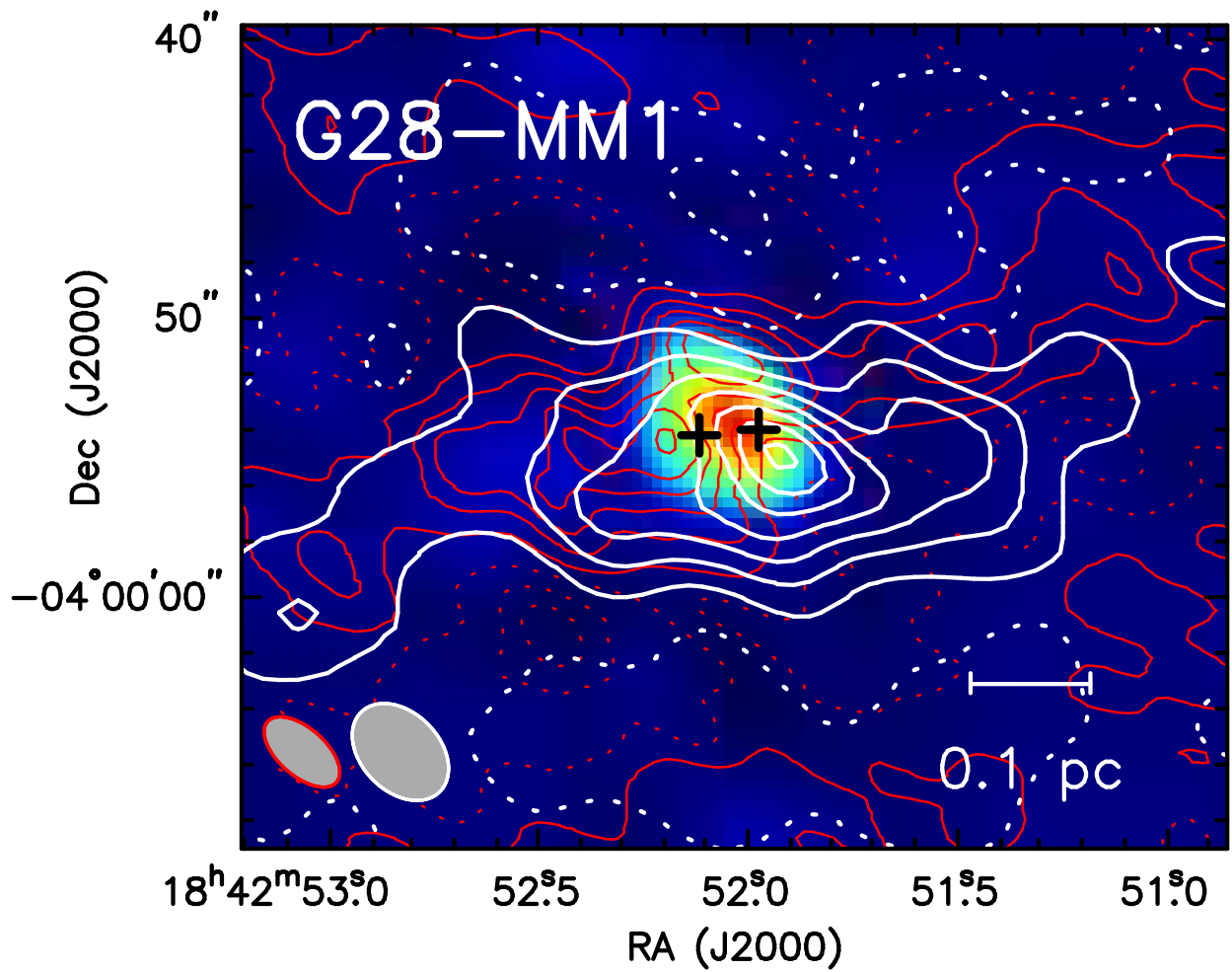} %
\caption{SMA $\mathrm{N_2H^+ \; (3-2)}$ integrated intensity ({\it red contours}) and $\mathrm{^{13}CO \; (2-1)}$ integrated intensity ({\it white contours}) overlaid with the line-free continuum map ({\it colors}) of MM1.  Contour levels correspond to $(-10, -6, -2, 2, 6, \dots, 22) \, \sigma$, where $\sigma = 1.10$ and $1.12 \; \mathrm{K \, km \, s^{-1}}$ for $\mathrm{N_2H^+}$ and CO, respectively.  
Black crosses indicate the continuum peak positions (Zhang {et~al.} 2009).  
The synthesized beam size is $4.0\arcsec \times 2.8\arcsec$ ($\mathrm{P.A.} = 44^\circ$) for CO.
\label{fig_g28mm1}} %
\end{figure}

In MM1, a deficiency of $\mathrm{N_2H^+}$ is observed in the location of the dust continuum emission as well as the $\mathrm{^{13}CO \; (2-1)}$ emission (Fig.~\ref{fig_g28mm1}) in a previous study (Zhang {et~al.} 2009).
Observationally, a centrally heated temperature structure from $16$ to $30 \; \mathrm{K}$ is derived  in MM1 over spatial scales of roughly $1$ to $0.1 \; \mathrm{pc}$ (Pillai {et~al.} 2006; Zhang {et~al.} 2009). 
Theoretically, a significant drop in $D_\mathrm{frac}$ is predicted across this temperature range (Roueff {et~al.} 2005), and an instant release of CO is expected due to a significant drop of the CO sublimation timescale from $10^8 \; \mathrm{yr}$ at $T_d \simeq 12 \; \mathrm{K}$ to $0.1 \; \mathrm{yr}$ at $T_d \simeq 20 \; \mathrm{K}$ (Collings {et~al.} 2003). 
This hypothesized release of CO in the warm region can be traced in MM1 with the $\mathrm{^{13}CO}$ emission and dust continuum emission.
A contrast in chemical composition can occur across the boundary between the cold, outer part and the warm, inner part that has been altered by the newly released gas-phase reactants.
This chemical contrast has recently been observed in AFGL~5142 by comparing emissions of $\mathrm{N_2H^+}$ and $\mathrm{NH_3}$ (Busquet {et~al.} 2009).

\placefigure{f3} %

\section{Summary} %
Our main findings are summarized as follows:
\begin{enumerate} %
\item A moderate decreasing trend over a factor of $3$ in deuterium fractionation of $\mathrm{N_2H^+}$ with evolutionary stage.
Such trend extends the use of the $N(\mathrm{N_2D^+})/N(\mathrm{N_2H^+})$ ratio as an evolutionary tracer to high-mass protostellar candidates.
\item An $\mathrm{N_2H^+}$ void resulting in part from a warmer temperature of the core and an instant release of CO from sublimated grain mantles in the warm region of MM1.    
\item A large fraction of the $\mathrm{N_2H^+} \; (3-2)$ flux missed by the interferometer suggesting the presence of cold and dense gas over a rather large area ($\gtrsim 0.2 \; \mathrm{pc}$) in G28.
\end{enumerate} %
 
\acknowledgments %
This research is supported by National Science Council of Taiwan through grants NSC 97-2112-M-001-006-MY2 and NSC 97-2112-M-007-006-MY3.

\clearpage
\begin{deluxetable}{lccc} %
\tablewidth{0pt} %
\tablecolumns{4} %
\tablecaption{Results of SMT Spectral Fits \label{table_smtfits}} %
\tablehead{ \colhead{Parameters} & \colhead{MM1} & \colhead{MM4} & \colhead{MM9} } %
\startdata
$T_\mathrm{ex} \; \mathrm{(K)}$ & 16.0 & 16.6 & 13.2 \\
\tableline
\sidehead{$\mathrm{N_2H^+ \; (3-2)}$:} 
$N$ ($10^{12} \; \mathrm{cm^{-2}}$) & $10.12 \pm 0.05$ & $5.14 \pm 0.04$ & $1.86 \pm 0.05$ \\
$\upsilon_\mathrm{LSR}$ ($\mathrm{km \, s^{-1}}$) & $77.59 \pm 0.02$ & $78.63 \pm 0.02$ & $79.82 \pm 0.04$ \\
$\Delta \upsilon$ ($\mathrm{km \, s^{-1}}$) & $6.14 \pm 0.04$ & $3.88 \pm 0.04$ & $3.42 \pm 0.11$ \\
$\overline{\chi^2}$ & 122.9 & 9.2 & 1.5 \\
$\tau_\mathrm{max}$ & 0.21 & 0.16 & 0.08 \\
\tableline
\sidehead{$\mathrm{N_2D^+ \; (3-2)}$:} 
$N$ ($10^{11} \; \mathrm{cm^{-2}}$) & $1.7 \pm 0.2$ & $1.2 \pm 0.3$ & $1.0 \pm 0.2 $ \\
$\upsilon_\mathrm{LSR}$ ($\mathrm{km \, s^{-1}}$) & $78.0 \pm 0.2$ & $79.6 \pm 0.4$ & $81.1 \pm 0.2$ \\
$\Delta \upsilon$ ($\mathrm{km \, s^{-1}}$) & $2.6 \pm 0.5$ & $3.7 \pm 0.9$ & $2.2 \pm 0.6$ \\
$\overline{\chi^2}$ & 1.8 & 0.9 & 1.4 \\
$\tau_\mathrm{max}$ & 0.007 & 0.003 & 0.005 \\
\tableline
$D_\mathrm{frac}$ & $0.017 \pm 0.002$ & $0.024 \pm 0.005$ & $0.052 \pm 0.011$ 
\enddata %
\end{deluxetable} %

\begin{deluxetable}{lccc} %
\tablewidth{0pt} %
\tablecolumns{4} %
\tablecaption{Comparison of Three Sources \label{table_mms}} %
\tablehead{ \colhead{Parameters} & \colhead{MM1} & \colhead{MM4} & \colhead{MM9} } %
\startdata %
\sidehead{$\mathrm{N_2H^+ \; (3-2)}$:} %
$W_\mathrm{SMA}$ ($\mathrm{K \, km \, s^{-1}}$) & 1.12 & 0.33 & 0.05 \\
$W_\mathrm{SMT}$ ($\mathrm{K \, km \, s^{-1}}$) & 12.93 & 7.17 & 2.23 \\
$W_\mathrm{SMA}/W_\mathrm{SMT}$ (\%) & 8.6 & 4.6 & 2.4 \\
\tableline
\sidehead{Continuum emission:} %
$T_\mathrm{ex}$ & 30.0 & 13.0 & 13.7 \\
$F_\mathrm{236.5 \; GHz}$ (Jy) & 0.59 & 0.15 & 0.03 \\ 
$F_\mathrm{284.2 \; GHz}$ (Jy) & 1.24 & 0.28 & 0.04 \\
$\beta$ & 2.3 & 2.0 & 0.9 \\
$M_\mathrm{core}$ ($M_\odot$) & 277 & 209 & 36  
\enddata %
%
\end{deluxetable} %

\end{document}